\documentstyle[11pt,newpasp,epsfig,twoside]{article}
\newcommand{\Lya}{Ly$\alpha\ $}

\newcommand{\etal}{et~al.\ }
\def\kms{\,{\rm km\,s^{-1}}}

\def\msun{\,{\rm M_\odot}}

\def\lya{Ly$\alpha\ $}

 \def\ni{\noindent}
\def\rei{{\rm rei}}
\def\obs{{\rm obs}}

\def\em{{\rm em}}

\def\HI{\hbox{H~$\scriptstyle\rm I\ $}}
\def\HII{\hbox{H~$\scriptstyle\rm II\ $}}
\def\bHII{\hbox{H~$\scriptstyle\rm \bf II\ $}}

\def\HeII{\hbox{He~$\scriptstyle\rm II\ $}}
\def\HeIII{\hbox{He~$\scriptstyle\rm III\ $}}

\def\nH{{\rm H}}

\def\ni{{\noindent}}

\begin{document}

\title{Probing the End of the Dark Ages}
\author{Piero Madau}
\affil{Department of Astronomy and Astrophysics, University of 
California, Santa Cruz, CA 95064, USA.}

\begin{abstract}
\ni In currently popular cosmological scenarios -- all variants of the cold 
dark matter (CDM) cosmogony -- some time beyond a redshift of 15,
stars within the numerous small halos that condense with virial 
temperatures $\sim 10^4\,$K created the first heavy elements; these 
protogalactic systems, together 
perhaps with an early population of mini-quasars, generated the ultraviolet 
radiation and mechanical energy that reheated and reionized the cosmos.
The history of the Universe during and soon after these crucial formative
stages is recorded in the thermal, ionization, and chemical state of the 
all-pervading intergalactic medium (IGM), which
contains most of the ordinary baryonic
material left over from the big bang. Throughout the epoch of structure
formation, the IGM becomes clumpy under the
influence of gravity, and acts as a source for the gas that gets
accreted, cools, and forms stars within subgalactic fragments, and as a sink 
for the metal enriched material, energy, and radiation which they eject.

\end{abstract}

\section{Introduction}
\ni At epochs corresponding to $z\sim 1000$ the intergalactic medium 
is expected to recombine and remain neutral until sources of radiation and 
heat develop that are capable of
reionizing it. The detection of transmitted flux shortward of the \lya
wavelength in the spectra of sources at $z\sim 6$ implies that the hydrogen
component of this IGM was ionized at even higher redshifts.
It appears then that substantial sources of ultraviolet photons and mechanical
energy were already
present when the Universe was less than 6\% of its current age, perhaps
quasars and/or young star-forming galaxies:
an episode of pregalactic star formation may provide a possible explanation
for the widespread existence of heavy elements (like carbon, oxygen, and
silicon) in the IGM, while the integrated radiation emitted from quasars is 
likely responsible for the reionization of intergalactic helium at later 
times.

Popular cosmological models predict that some time beyond a redshift of
15 the gas
within halos with virial temperatures $T_v\ga 10^4\,$K [or, equivalently,
with masses $M\ga 10^9\, (1+z)^{-3/2}h^{-1}\,\msun$ comparable to present-day 
dwarf ellipticals] cooled rapidly due to the excitation of
hydrogen \Lya by the Maxwellian tail of the electron distribution, and
fragmented. Massive stars formed with some initial mass function (IMF),
synthesized heavy elements, and exploded as
Type II supernovae (SNe) after a few $\times 10^7\,$yr, enriching and heating 
the surrounding medium. Whilst collisional excitation
of molecular hydrogen may have allowed the gas in even smaller systems [virial
temperatures of only a few hundred K, corresponding to masses around
$10^7\,(1+z)^{-3/2}h^{-1}\,\msun$] to cool and form stars at earlier times
(Abel, Ciardi, and Ferrara, this volume), H$_2$ molecules are
efficiently photo-dissociated by stellar UV radiation, and such negative
`feedback' is likely to have suppressed molecular cooling and further star
formation inside very small halos.
One should note that while numerical N-body$+$hydrodynamical simulations
have convincingly shown that the IGM is expected to fragment into structures at
early times in CDM cosmogonies (e.g. Cen \etal 1994; Zhang \etal 1995),
the same simulations are much less able to predict the efficiency with
which the first gravitationally collapsed objects lit up the Universe
at the end of the `dark age'.

\section{`Minihalos', Metal Enrichment, and Reionization}

\ni The reionization scenario that has been the subject of the most theoretical 
study is one in which intergalactic hydrogen is photoionized by the UV 
radiation emitted either by quasars or
by stars with masses $\ga 10\,\msun$, rather than ionized by collisions with 
electrons heated up by, e.g., supernova-driven winds from early pregalactic
objects. In the former case a high degree of ionization requires
about $13.6\times (1+t/\overline t_{\rm rec})\,$eV per hydrogen atom, where 
$\overline t_{\rm rec}$ is the volume-averaged hydrogen recombination 
timescale, the ratio $t/\overline t_{\rm rec}$ being much greater than unity 
already at $z\sim 10$ (Haiman and Gnedin, this volume). 
Collisional ionization to a neutral fraction of only
few parts in $10^{5}$ requires a comparable energy input, i.e. an IGM 
temperature close to $10^5\,$K or about $25\,$eV per atom. 

Massive stars will deposit both radiative and mechanical energy into the 
interstellar medium of subgalactic fragments. A complex network of `feedback' 
mechanisms is likely at work in these systems, as the gas in shallow potential
is more easily blown away thereby quenching further star formation, and the 
blastwaves produced by supernova explosions reheat 
the surrounding intergalactic gas and enrich it with newly formed heavy 
elements and dust. It is therefore unclear at this stage whether an early 
input of mechanical energy will actually play a major role in determining the 
thermal and ionization state of the IGM on large scales (Tegmark \etal 1993). 
What can be easily shown is that, during the evolution of a a 
`typical' 
stellar population, more energy is lost in ultraviolet radiation than in
mechanical form. This is because in nuclear burning from zero to solar 
metallicity ($Z_\odot=0.02$), the energy radiated per 
baryon is $0.02\times 0.007\times m_\nH c^2$; about one third of it goes into 
H-ionizing photons. The same massive stars that dominate the UV light 
also explode as SNe, returning most of the metals to the 
interstellar medium and 
injecting about $10^{51}\,$ergs per event in kinetic energy. For a Salpeter 
IMF, one has about one SN every $150\,\msun$ of baryons
that forms stars. The mass fraction in mechanical energy is then approximately
$4\times 10^{-6}$, ten times lower than the fraction released in photons
above 1 ryd. 

The relative importance of photoionization versus shock ionization will 
depend, however, on the efficiency with which radiation and mechanical 
energy actually escape into the IGM.    
Consider, in particular, the case of an early generation of halos with circular 
speed $v_c=25\,\kms$, corresponding in top-hat spherical collapse to a virial 
temperature $T_v=0.5\mu m_p v_c^2/k\approx 10^{4.3}\,$K
and halo mass $M=0.1v_c^3/GH\approx 10^8 [(1+z)/10]^{-3/2}
h^{-1}\, \msun$. In these systems rapid cooling by
atomic hydrogen can take place and a significant fraction,
$f\Omega_B$, of their total mass may be converted into stars over a 
dynamical timescale (here $\Omega_B$ is the baryon density parameter). 
In a flat cosmology with $\Omega_M=1$, $h=0.5$, and rms 
mass fluctuation
normalized at present to $\sigma_8=0.63$ on spheres of $8\,h^{-1}\,$Mpc,
halos with $M=10^8\,h^{-1}\,\msun$ would be collapsing at $z=9$
from 2 $\sigma$ fluctuations. At this epoch more massive halos with
$M=10^{10}\,h^{-1}\,\msun$, while able to cool rapidly, would be collapsing 
from 3 $\sigma$ peaks and be too rare to produce significant amounts of heavy 
elements and UV radiation (in a gaussian theory the 3 $\sigma$ peaks contain 
only 5\% as much mass as the 2 $\sigma$ peaks at a given epoch), unless they 
were able somehow to form stars more efficiently than lower-mass objects. 
Halos from 1 $\sigma$ fluctuations would be more numerous and contain most of 
the mass, but with virial temperatures of only a few hundred degrees they 
would most likely be unable to cool via H$_2$ before reionization actually
occurs at $z=z_\rei$. 

For $f=0.1$, $\Omega_Bh^2=0.019$, and $h=0.5$, the 
explosive output of $10,000$ SNe would inject a total energy $E_0\approx 
10^{55}\,$ergs. Correlated multi-SN explosions will create large holes in 
the ISM of these pregalactic halos, perhaps
enlarging by far pre-existing ones due to winds from their progenitors
stars. The energy from SNe will impart enough momentum to the interstellar 
medium that large portions of it will become unbound 
and leave the parent halo, taking the metal-enriched stellar debris along  
(MacLow \& Ferrara 1999).
A large fraction -- perhaps as much as 90\% -- of the energy injected in these 
SN-driven bubbles could be lost to radiation in the halo before the bubbles 
expand out into the surrounding IGM (Madau \etal 2000).
Eventually the hot gas will
escape its host, shock the IGM, and form a cosmological blast wave.
If the explosion occurs at cosmic time $t=4\times 10^8\,$yr (corresponding in 
the adopted cosmology to $z=9$), at time $\Delta t=0.2\,t$ 
after the event we can use the standard Sedov-Taylor self-similar (adiabatic)
solution to give a rough estimate of the proper radius of the shock,
$$ 
R_s\approx \left[\frac{12\pi G (0.1 E_0)}{\Omega_B}\right]^{1/5}t^{2/5}\Delta 
t^{2/5} \approx 10\,{\rm kpc}.  \eqno(1)  
$$
At this instant the shock velocity relative to the Hubble flow is  
$$
v_s\approx 2R_s/5\Delta t\approx 50\,\kms,  \eqno(2)  
$$
lower than the escape velocity from the halo. The gas temperature 
just behind the shock front is $T_s=3\mu m_pv_s^2/16k 
\approx 10^5\,$K, more than enough to efficiently ionize all the 
incoming hydrogen. At these redshifts, it is the onset of Compton cooling off 
cosmic microwave background photons that ends the adiabatic stage of blast 
wave propagation. According to the Press-Schechter formalism, the 
comoving abundance of collapsed halos with mass $M=10^8\,h^{-1}\,M_\odot$ 
at $z=9$ is $dn/d\ln M\sim 30\,h^3\,$Mpc$^{-3}$, corresponding to a mean 
proper distance between neighboring halos of $\sim 20\,h^{-1}\,$kpc, and to a 
total mass density parameter of order $0.01$. With the 
assumed star formation efficiency only a small fraction, about one 
percent, of the stars seen today would form at these early epochs. Still, our 
simplistic analysis (see Madau \etal 2000 for a more detailed 
treatment) conveys the interesting idea that the blast waves from such 
a population of pregalactic objects may actually fill a significant fraction 
of the Hubble volume, and drive 
the intergalalactic medium at early epochs to a higher adiabat, $T\sim 
10^5\,$K, than expected from photoionization, so as to inhibit the formation
of further protogalaxies by raising the Jeans mass. This effect could perhaps
be responsible for providing a global negative feedback to self-regulate the
early stellar birthrate, effectively causing a pause in the cosmic 
history of star formation. After this epoch of pregalactic outflows, the
IGM would be polluted to a mean metallicity $\langle Z\rangle=\Omega_Z/
\Omega_b\sim 0.001\,Z_\odot$ (Madau \etal 2000).
A lower density of sources -- which 
would therefore have to originate from higher amplitude peaks -- would suffice 
if the typical efficiency of star formation were larger than assumed here. 

\section{Cosmological \bHII regions}

\ni In this section we will focus our attention to the  
photoionization of the IGM, i.e. we will assume that UV
photons from an early generation of stars and/or quasars are the main source
of energy for the reionization and reheating of the Universe, and that 
star formation and quasar activity occurs in collapsed galaxy halos.  
The process then begins as individual sources start to generate
expanding \HII regions in the surrounding IGM; throughout an \HII region, H is
ionized and He is either singly or doubly ionized. As more and more sources of
ultraviolet radiation switch on, the ionized volume grows in size while the 
neutral phase shrinks. Reionization is completed when the \HII regions 
overlap, and every point in the intergalactic space gets exposed for the first 
time to a nearly uniform Lyman-continuum (Lyc) background.

In the presence of a population of ionizing sources, the transition from a
neutral IGM to one that is almost fully ionized can be statistically described 
by the evolution with redshift of the volume filling factor (or porosity)
$Q$ of \HII, \HeII, and \HeIII regions. The radiation emitted by spatially
clustered stellar-like and quasar-like sources -- the number densities and
luminosities of which may change rapidly as a function of redshift --
coupled with absorption processes in a medium that becomes more and more 
clumpy owing to the non-linear collapse of structures (Figure 1), all
determine the complex topology of neutral and ionized zones in the Universe
(Gnedin, this volume; Ciardi \etal 2000; Abel \etal 1999).
When $Q\ll1$ and the radiation sources are randomly distributed, the ionized
regions are spatially isolated, every UV photon is absorbed somewhere in the
IGM, and the UV radiation field is highly inhomogeneous. As $Q$ grows, the 
crossing of ionization fronts becomes more and more common, until 
percolation occurs at $Q=1$.

\begin{figure}
\epsfysize=5cm 
\epsfxsize=5cm 
\hspace{3.5cm}\epsfbox{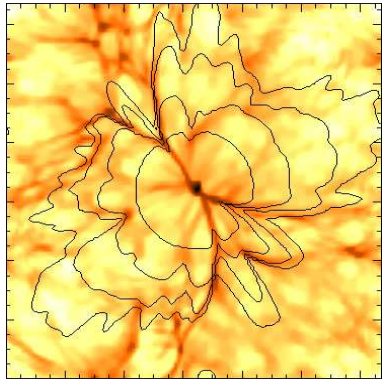}
\vspace{0.5cm}
\caption{\small Simulating the reionization of the Universe:
propagation of an ionization front in a $128^3$
cosmological density field. A `mini-quasar' emitting $5\times
10^{53}\,$ ionizing photons s$^{-1}$ was turned on at the densest cell, in 
a virialized halo
of total mass $10^{11}\,M_\odot$. The box length is 2.4 comoving
Mpc. The solid contours give the position of the front at 0.15, 0.25, 0.38, and
0.57\,Myr after the quasar has switched on at $z=7$. The underlying greyscale
image indicates the initial \HI density field. (From Abel \etal 1999.)
}
\end{figure}

Since the mean free path of Lyc radiation is always much 
smaller than the horizon (this is also true after `overlapping' because of 
the residual \HI still present in the \lya forest clouds and the Lyman-limit 
systems), the filling factor of cosmological \HII regions is 
equal at any given time $t$ to the total number of ionizing photons 
emitted per hydrogen atom by all radiation sources present at earlier epochs,
$\int_0^t \dot n_{\rm ion}dt'/\overline{n}_\nH$, minus the total number 
of radiative
recombinations per atom, $\int_0^t Q dt'/\overline{t}_{\rm rec}$. This
statement reflects the simple fact that every ultraviolet photon that is
emitted is either absorbed by a newly ionized hydrogen atom or by a 
recombining one. Differentiating one gets
$$
\frac{dQ}{dt}=\frac{\dot n_{\rm ion}}{\overline{n}_\nH}-\frac{Q}
{\overline{t}_{\rm rec}}.  \eqno(3)
$$
It is this differential equation -- and its equivalent for
expanding helium zones -- that statistically describes the transition
from a neutral Universe to a fully ionized one (Madau \etal 1999). 
Initially, when the filling factor is 
$\ll 1$, recombinations can be neglected and the ionized volume increases 
at a rate fixed solely by the ratio $\dot n_{\rm ion}/\overline{n}_\nH$. 
As time goes on and more and more Lyc photons are emitted, radiative 
recombinations become important and slow down the growth of the ionized 
volume, until $Q$ reaches unity, the recombination term saturates, and 
reionization is finally completed (except for the high density regions 
far from any source which are only gradually eaten away, Miralda-Escud\'{e}, 
Haehnelt, \& Rees 2000). In the limit of a fast recombining IGM 
($\overline{t}_{\rm rec}\ll t$), one can neglect the derivative on the 
left-hand side of equation (3) and derive   
$$
Q\la \frac{\dot n_{\rm ion}}{\overline{n}_\nH}\overline{t}_{\rm rec},
\eqno(4)
$$
i.e. the volume filling factor of ionized bubbles must be less (or equal) to the
number of Lyc photons emitted per hydrogen atom in one recombination time. 
In other words, because of radiative recombinations, only a fraction 
$\overline{t}_{\rm rec}/t\ll 1$ of the photons emitted above 1 ryd is actually used 
to ionize new IGM material. The Universe is completely reionized when 
$$
\dot n_{\rm ion}\overline{t}_{\rm rec}\ga \overline{n}_\nH,
\eqno(5)
$$
i.e. when the emission rate of ultraviolet photons exceeds the mean rate of
recombinations.
 
\section{A clumpy Universe}

\ni The simplest way to treat reionization in a inhomogeneous medium is in 
terms of a clumping factor that increases the effective gas recombination rate. 
In this case the volume-averaged recombination time is
$$
\overline{t}_{\rm rec}=[(1+2\chi) \overline{n}_p\alpha_B\,C]^{-1}=0.06\, 
{\rm Gyr}
\left(\frac{\Omega_B h^2}{0.019}\right)^{-1}\left(\frac{1+z}{10}\right)^{-3}
\frac{\overline n_\nH}{\overline n_p}~C_{10}^{-1}, \eqno(6)
$$
where $\alpha_B$ is the recombination coefficient to the excited states of 
hydrogen (at an assumed gas temperature of $10^4\,$K), $\chi$ the helium to 
hydrogen abundance ratio, and the factor $C\equiv \langle n_p^2\rangle/
\overline{n}_p^2>1$ takes into account the degree of clumpiness of photoionized 
regions (hereafter $C_{10}\equiv C/10$). If ionized gas with density $n_p$ 
filled uniformly a fraction $1/C$ 
of the available volume, the rest being empty space, the mean square density 
would be $\langle n_p^2\rangle =n_p^2/C=\overline{n}_p^2C$. More in general, if 
$f_m$ is the fraction of baryonic mass in photoionized 
gas at an overdensity $\delta$ relative to the mean, and the remaining 
(underdense) medium is distributed uniformly, then the fractional volume 
occupied by the denser component is   
$$
f_v=f_m/\delta, \eqno(7) 
$$
the density of the diffuse component is
$$
\overline{n}_p\frac{1-f_m}{1-f_v}, \eqno(8) 
$$
and the recombination rate is larger than that of a homogeneous Universe by
the factor 
$$
C=f_m\delta+\frac{(1-f_m)^2}{1-f_v}. \eqno(9)
$$  
It is difficult to estimate the clumping factor accurately. According to 
hydrodynamics simulations of structure formation in the IGM (within the 
framework of 
CDM-dominated cosmologies), \Lya forest clouds with moderate overdensities, 
$5\la\delta\la 10$, occupy a fraction of the available volume which
is too small for them to dominate the clumping at high
redshifts (e.g. Theuns \etal 1998). In hierarchical 
clustering models, it is the virialized gas with $\delta\approx 200$ in the
earliest non-linear objects with $T_J\le T_v\le 10^{4}\,$K (here $T_J$ is the 
virial temperature corresponding to the cosmological Jean mass) which will 
boost the recombination rate by large factors as soon as the 
collapsed mass fraction exceeds $\sim 0.5$\% (Haiman, this volume). 
The importance of photoevaporating minihalos as 
sinks of ionizing photons during cosmological reionization has been recently
discussed by Haiman \etal (2000). Note that such minihalos are not 
yet resolved in large-scale three-dimensional cosmological simulations. 
Halos or halo cores which are dense and thick enough to be self-shielded 
from UV radiation will stay neutral 
and will not contribute to the recombination rate. This is also true of gas 
in more massive halos, which will be virialized to higher temperatures 
and ionized by collisions with thermal electrons. 

\section{The Earliest Luminous Sources and the Damping Wing of the Gunn-Peterson
Trough}

\ni Prior to complete reionization, sources of
ultraviolet radiation will be seen behind a large column of intervening gas 
that is still neutral. In this case, because of scattering off the 
line-of-sight due to the diffuse neutral IGM, the spectrum of a source at 
$z_\em>z_\rei$ should show the red damping wing of the Gunn-Peterson 
absorption trough at wavelengths longer 
than the local \Lya resonance, $\lambda_\obs>\lambda_\alpha(1+z_\em)$, where
$\lambda_\alpha=c/\nu_\alpha=1216\,$A.
At $z\ga 6$, this characteristic feature extends for more than $1500\,\kms$ 
to the red of the resonance, and may significantly suppress the \Lya 
emission line in the spectra of the first generation of objects in the
Universe. Measuring the shape of the absorption profile of the damping wing
could provide a determination of the density of the neutral IGM near the 
source (Miralda-Escud\'{e} 1998). 

\begin{figure}
\epsfysize=7cm 
\epsfxsize=7cm 
\hspace{3.5cm}\epsfbox{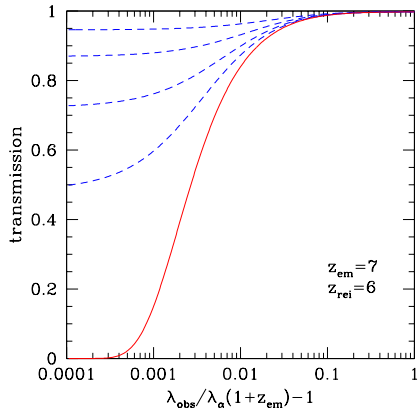}
\vspace{0.0cm}
\caption{\small The red damping wing of the Gunn-Peterson trough for a source at $z_\em
=7$. The transmission is plotted as a function of the
fractional wavelength interval from the \Lya resonance at $\lambda_\alpha
(1+z_\em)$. The neutral IGM is
assumed to be completely reionized by $z_\rei=6$.
{\it Solid curve:} absorption profile neglecting the effect of the local \HII
region around the source. Note how a fraction $\ga \exp(-0.5)=0.6$ of
the radiated flux will only be transmitted $\ga 1200\,\kms$ (corresponding to
$\ga 40\,$A) to the red of the resonance.
{\it Dashed curves:} absorption profiles including the local
\HII zone generated by a QSO shining for $10^7\,$yr. From top to bottom, the
curves are plotted for four decreasing ionizing photon luminosities,
$\dot N_i=10^{58}$, $10^{57},$ $10^{56}\,$, and $10^{55}\,$s$^{-1}$
(Einstein-de Sitter Universe with $h=0.5$ and $\Omega_Bh^2=0.019$).
}
\end{figure}

Cen \& Haiman (2000) and Madau \& Rees (2000) have recently focused on the 
width of the red damping wing -- related to the expected strength of the \Lya 
emission line -- in the spectra of very distant QSOs as a flag of the 
observation of the IGM before reionization. 
They have assessed, in particular, the impact of the photoionized, Mpc-size
regions which will surround individual luminous sources of Lyc radiation
on the transmission of photons redward of the \Lya resonance, and shown that 
the damping wing of the Gunn-Peterson trough may nearly completely disappear
because of the lack of neutral hydrogen in the vicinity of a bright QSO.
If the quasar lifetime
is shorter than the expansion and gas recombination timescales, the volume
ionized will be proportional to the total number of photons produced
above 13.6 eV: the effect of this local photoionization is to greatly
reduce the scattering opacity between the redshift of the quasar
and the boundary of its \HII region. The 
transmission on the red side of the \Lya resonance is always greater than 
50\% for sources radiating a total of
$\ga 10^{69.5}$ ionizing photons into the IGM (Figure 2). The detection of 
a strong \Lya
emission line in the spectra of bright QSOs shining for $\ga 10^7\,$yr
cannot then be used, by itself, as a constraint on the reionization epoch. The
first signs of an object radiating prior to the transition from a neutral to
an ionized Universe may be best searched for in the spectra of luminous sources
with a small escape fraction of Lyman-continuum photons into the IGM, or sources
with a short duty cycle.

\acknowledgments
\ni I would like to thank my collaborators, T. Abel, A. Ferrara, F. Haardt, Z. 
Haiman, and M. Rees, for many useful discussions on the topics discussed here.

\end{document}